\pdfoutput=1
\documentclass[prb,twocolumn,showpacs,amsmath,amssymb,superscriptaddress,floatfix]{revtex4-1}

\usepackage{graphicx}
\usepackage{dcolumn}
\usepackage{bm}

\begin{document}

\title{Magnetoresistance and negative differential resistance in Ni$|$Graphene$|$Ni vertical heterostructures driven by finite bias voltage: A first-principles study}

\author{Kamal K. Saha}
\affiliation{Department of Physics and Astronomy, University of Delaware, Newark, DE 19716-2570, USA}
\author{Anders Blom}
\affiliation{QuantumWise A/S, Lers{\o} Parkall\'{e} 107, 2100 Copenhagen, Denmark}
\author{Kristian S. Thygesen}
\affiliation{Center for Atomic-scale Materials Design (CAMD), Department of Physics, Technical University of Denmark, DK-2800 Kongens Lyngby, Denmark}
\author{Branislav K. Nikoli\' c}
\email{bnikolic@udel.edu}
\affiliation{Department of Physics and Astronomy, University of Delaware, Newark, DE 19716-2570, USA}

\begin{abstract}
Using the nonequilibrium Green function formalism combined with density functional theory, we study finite-bias quantum transport in  Ni$|$Gr$_n|$Ni vertical heterostructures where $n$ graphene layers are sandwiched between two semi-infinite Ni(111) electrodes. We find that recently predicted ``pessimistic'' magnetoresistance of 100\%  for $n \ge 5$ junctions at zero bias voltage \mbox{$V_b \rightarrow 0$}, persists up to \mbox{$V_b \simeq 0.4$ V}, which makes such devices promising for spin-torque-based device applications. In addition, for parallel orientations of the Ni magnetizations, the $n=5$ junction exhibits a pronounced negative differential resistance as the bias voltage is increased from \mbox{$V_b=0$ V} to \mbox{$V_b \simeq 0.5$ V}. We confirm that both of these nonequilibrium transport effects hold for different types of bonding of Gr on the Ni(111) surface while maintaining Bernal stacking between individual Gr layers.
\end{abstract}


\maketitle

\section{Introduction}\label{sec:intro}

A magnetic tunnel junction (MTJ) consists of an ultrathin insulating barrier which separates two metallic ferromagnetic (F) layers with variable magnetization direction. The MTJs based on transition metals or their alloys and an epitaxial MgO barrier~\cite{Butler2001} are the present workhorse of both commercial and basic research spintronics. For example, MgO-based MTJs are the core elements of read heads in hard drives or in magnetic random-access memory devices~\cite{Katine2008} that are operated by the current-induced spin-transfer torque (STT). In the STT phenomenon,
spin current of large enough density injected into a free F layer either switches its magnetization from one static configuration to another or generates a dynamical situation with steady-state precessing magnetization.~\cite{Ralph2008} Thus, the usage of MTJs in STT-based spintronic devices
necessitates~\cite{Katine2008,Wang2011} a compromise between large current density (which requires low junction resistance to avoid damage) driven by finite bias voltage and readability (which requires large magnetoresistance).

One of the great successes of first-principles quantum transport modeling has been a conjecture~\cite{Butler2001} of {\em very large} ``optimistic''  tunneling magnetoresistance, $\mathrm{TMR} = (G_{\rm P}-G_{\rm AP})/G_{\rm AP} \times 100\% \gtrsim 1000\%$, in Fe$|$MgO$|$Fe(100) MTJs where $G_{\rm P}$ ($G_{\rm AP}$) is conductance for parallel (antiparallel) orientation of the Fe magnetizations. This prediction has ignited intense fabrication efforts reaching TMR of about 200 \% at room temperature~\cite{Yuasa2004} which, although undoubtedly correlated with the crystallinity of MgO barrier, is difficult to reconcile with first-principles predictions.~\cite{Butler2001} The origin of the discrepancy is the sensitivity of spin injection and TMR to details of difficult-to-control interfacial disorder as revealed by a number of theoretically investigated scenarios (such as the intermixing of Fe and MgO,~\cite{Mathon2006} oxygen vacancies at or near the Fe$|$MgO interface,~\cite{Ke2010} or substoichiometric FeO layers with small oxygen concentrations~\cite{Bose2008}). In addition, TMR in MgO-based MTJs decays precipitously~\cite{Yuasa2004,Waldron2007,Rungger2009} with increasing bias voltage where the specific features of the decay are also sensitive to the type of interfacial disorder.~\cite{Ke2010}

\begin{figure}
\includegraphics[scale=0.35,angle=0]{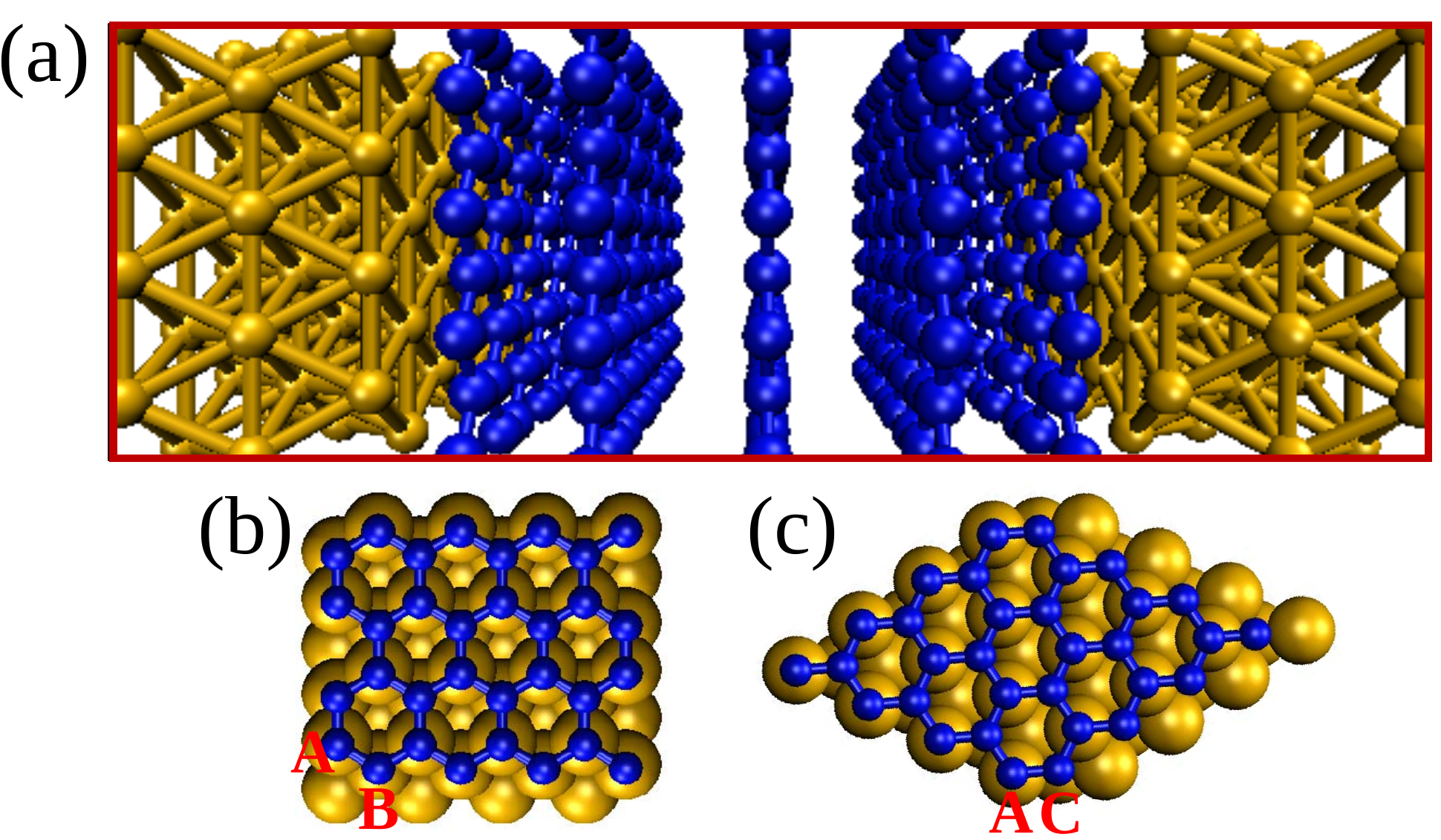}
\caption{(a) Schematic view of Ni$|$Gr$_5|$Ni junction where Gr$_5$
represents five layers of graphene and Ni is (111) fcc nickel. The device extends to infinity along the transverse directions while Ni electrode (orange) are semi-infinite in the longitudinal (transport) direction. The two investigated types of bonding~\cite{Karpan2007,Karpan2008,Fuentes-Cabrera2008} for Gr on the Ni(111) surface are illustrated in panel (b), as AB configuration where the two carbon atoms in the graphene unit cell cover Ni atoms in layers A (surface) and B (second layer), and panel (c) as AC configuration in which carbon atoms are placed directly above the Ni atoms in layers A (surface) and C (third layer). Here ABC refers to three close-packed layers within a fcc crystal.}
\label{fig:fig1}
\end{figure}

These issues could be resolved by searching for new material systems which would ensure perfect spin filtering in the absence of disorder while being much less sensitive to the presence of interfacial disorder in realistic junctions. For example, the recent first-principles analysis~\cite{Karpan2007,Karpan2008}  has brought an example of such system---Ni$|$Gr$_n|$Ni junctions---where $n$ layers of graphene (Gr$_n$) are sandwiched between two Ni electrodes as illustrated in Fig.~\ref{fig:fig1}. Graphene is recently discovered~\cite{Castro-Neto2009} two-dimensional (2D) allotrope of carbon where electronic states of a single layer Gr$_1$ or multilayers Gr$_n$ close to the Fermi energy are located around the high symmetry $K$ point in reciprocal space. The Ni$|$Gr$_n|$Ni junction exploits very small mismatch of $1.3$\% between the in-plane lattice constant of Gr and the surface lattice constant of Ni(111), as well as the fact that majority spin states of Ni are absent in a large region around the $K$ point. These two features combined lead to perfect spin filtering for $n \ge 5$, as  quantified by the ``pessimistic'' magnetoresistance $\mathrm{MR} = (G_{\rm P}-G_{\rm AP})/G_{\rm P} \times 100\% \approx 100$ \% (the ``optimistic'' MR diverges since $G_{\rm AP}$ vanishes for large $n$). The three times smaller lateral lattice mismatch compared to the 3.8\% for conventional Fe$|$MgO$|$Fe junctions should also reduce some of the strain and amount of defects that otherwise limit the thickness and degrade the efficiency of spin injection in MgO-based MTJs.

However, very little is known about {\em nonequilibrium} transport driven by finite bias voltage $V_b$ in Ni$|$Gr$_n|$Ni  junctions. This is partly due to the fact that standard first-principles electronic transport tools employed to capture electronic and magnetic structure at interfaces, such as layer Korringa-Kohn-Rostoker approach applied~\cite{Butler2001} to MgO-based MTJs or tight-binding muffin tin orbital wave-function matching scheme applied~\cite{Karpan2007,Karpan2008} to Ni$|$Gr$_n|$Ni junctions, become very cumbersome~\cite{Zhang2004b} to use at finite $V_b$ where one has to compute the charge redistribution~\cite{Areshkin2010} due to current flow by evaluating the nonequilibrium density matrix ${\bm \rho}$---a procedure which ensures the gauge invariance~\cite{Christen1996} of the current-voltage {\em I-V} characteristics. The nonequilibrium Green function formalism combined with density functional theory (NEGF-DFT),~\cite{Areshkin2010,Taylor2001,Brandbyge2002} where DFT part of the calculations is implemented in the basis of local orbitals, makes it relatively straightforward to obtain ${\bm \rho}$.

\begin{figure}
\includegraphics[scale=0.35,angle=0]{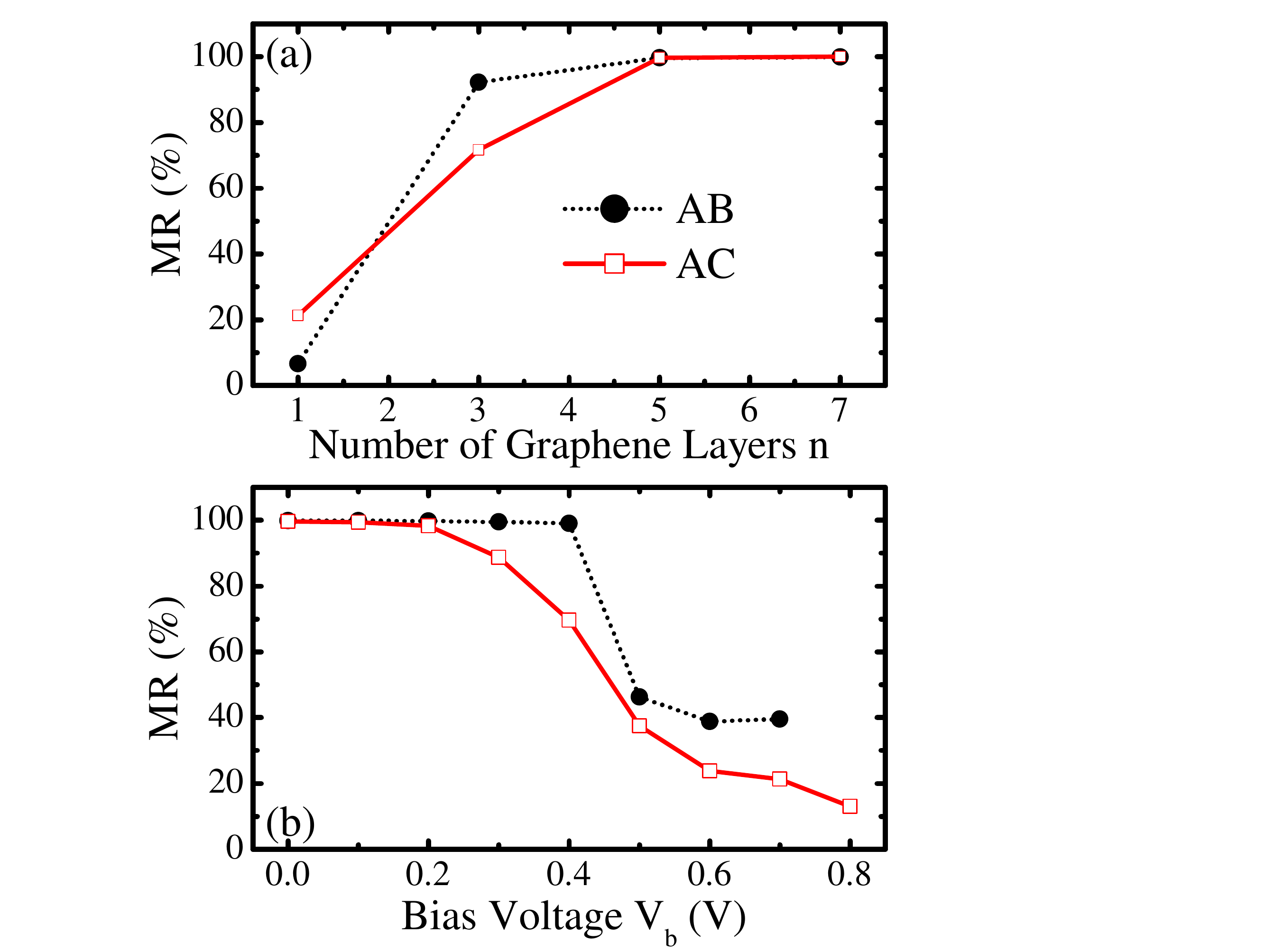}
\caption{(a) The ``pessimistic'' TMR for Ni$|$Gr$_n|$Ni junctions as a function of the number of graphene layers $n$ and for two different, AB and AC, bonding configurations for Gr on the Ni(111) surface illustrated in Figs.~\ref{fig:fig1}(b) and \ref{fig:fig1}(c), respectively. (b) The ``pessimistic'' TMR for $n=5$ junction versus finite bias voltage for AC and AB bonding configurations.}
\label{fig:fig2}
\end{figure}

Here we show how to use efficiently  spin- and ${\bf k}_\parallel$-resolved NEGF-DFT framework  to understand {\em nonequilibrium} transport through to Ni$|$Gr$_n|$Ni junctions depicted in Fig.~\ref{fig:fig1} for parallel (P) or antiparallel (AP) orientation of the Ni magnetizations. Our principal results are shown in Fig.~\ref{fig:fig2} and Fig.~\ref{fig:fig3}. In Fig.~\ref{fig:fig2}(a), we first confirm the result of Ref.~\onlinecite{Karpan2007,Karpan2008} about the zero bias ``pessimistic'' MR reaching 100\% for barriers composed of $n \ge 5$ graphene layers and, moreover, in Fig.~\ref{fig:fig2}(b) we predict that such maximized MR would persist even at finite $V_b \lesssim 0.4$ V. Figure~\ref{fig:fig2}(b) also suggests that bias voltage dependence of MR can be employed experimentally to determine the type of bonding configuration [illustrated in Figs.~\ref{fig:fig1}(b) and (c)] for Gr on the Ni(111) surface.

Furthermore, Fig.~\ref{fig:fig3}(c) shows that Ni$|$Gr$_n|$Ni junction with P orientation of the Ni magnetizations will exhibit pronounced negative differential resistance (NDR), where total charge current first increases and then decreases as the bias voltage is increased from $V_b=0$ V to $V_b \simeq 0.5$ V (or symmetrically in the opposite direction). The origin of NDR is explained in Fig.~\ref{fig:fig5} by plotting the position-dependent local density of states (LDOS) across the junction.

The paper is organized as follows. In Sec.~\ref{sec:negfdft} we discuss vertical Ni$|$Gr$_n|$Ni heterostructure setup and how to tune the spin- and ${\bf k}_\parallel$-resolved NEGF-DFT framework in order to describe properly magnetism around its interfaces. Section~\ref{sec:mr} discusses magnetoresistance at finite bias voltage, as well as the unusual conduction properties of general vertical graphene heterostructures, whose fabrication has been initiated recently,~\cite{Lee2011} which make Ni$|$Gr$_n|$Ni junctions different from  either conventional MTJs or spin valves. In Sec.~\ref{sec:mr}, we discuss NDR in Ni$|$Gr$_n|$Ni junctions driven by finite bias voltage. We conclude in Sec.~\ref{sec:conclusion}.

\section{The vertical heterostructure setup and tuning of NEGF-DFT framework for its modeling}\label{sec:negfdft}

The disorder-free junction shown in Fig.~\ref{fig:fig1}(a) consists of up to seven graphene layers arranged in Bernal stacking~\cite{Castro-Neto2009}  which serve as the barrier separating the two semi-infinite Ni electrodes. The junction is infinite in the transverse direction, so that its transverse periodicity requires $k$-point sampling.~\cite{Waldron2007} The spin injection and spin filtering in ferromagnetic multilayers
depends not only on the properties of the F electrodes but also on geometry, bonding and electronic and magnetic structure of the contact region, as emphasized by the studies~\cite{Butler2001,Waldron2007} of MgO-based MTJs. Therefore, we consider two different Gr on the Ni(111) surface bonding configurations illustrated in Figs.~\ref{fig:fig1}(b) and \ref{fig:fig1}(c).

\begin{figure}
\includegraphics[scale=0.35,angle=0]{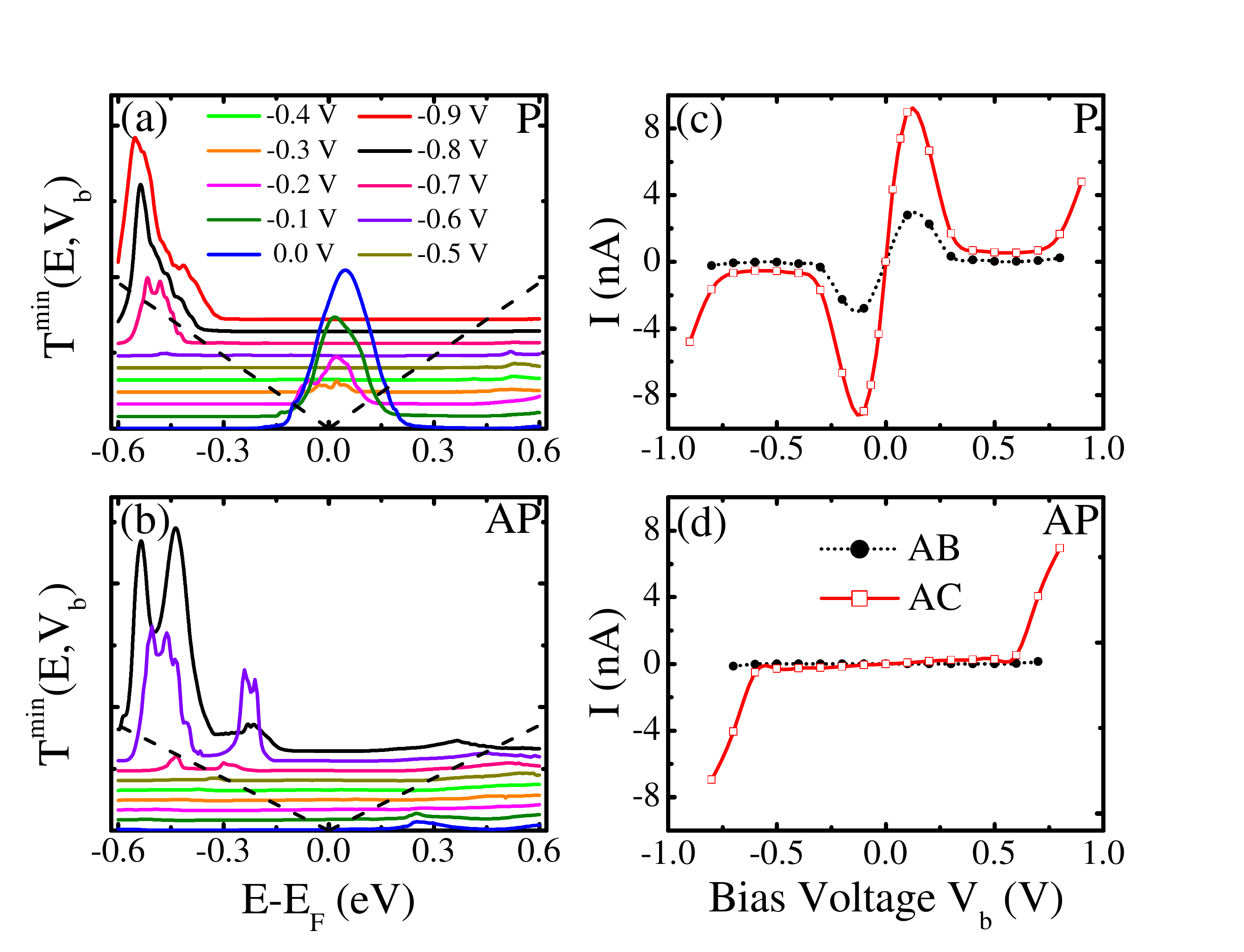}
\caption{The finite-bias transmission function $T^{\rm min}(E,V_b)$ for Ni$|$Gr$_n|$Ni junction in AC bonding configuration at the Ni(111)$|$Gr interface [Fig.~\ref{fig:fig1}(c)] for (a) P and (b) AP orientations of the Ni magnetizations. Since in P orientation minority spin contribution dominates, while in AP setup both minority and majority spins contribute the same, only $T^{\rm min}(E,V_b)$ is presented here for both P and AP orientations  with curves at different $V_b$ shifted along the $y$-axis for clarity. Panels (c) and (d) show {\em I-V} characteristics for P and AP orientation, respectively. The NDR is conspicuous in P orientation in panel (c) for both AC and AB bonding configurations.}
\label{fig:fig3}
\end{figure}

We note that DFT calculations employing different approximations for the exchange-correlation functional  (such as local density approximation,~\cite{Karpan2007,Karpan2008} generalized gradient approximation~\cite{Fuentes-Cabrera2008} and van der Waals density functional~\cite{Vanin2010,Mittendorfer2011}) have yielded contradictory conclusions about the AC bonding configuration being the most stable energetically and  the corresponding binding distance. The recent random phase approximation (RPA) calculations~\cite{Olsen2011} have resolved this controversy and demonstrated the conflicting results are due to a delicate interplay between covalent and dispersive interactions which is not captured by the DFT functionals. Also, the scanning tunneling microscopy imaging~\cite{Dedkov2010} shows that perfectly ordered epitaxial graphene layers can be prepared by elevated temperature decomposition of hydrocarbons where domains are larger than the terraces of the underlying Ni(111) surface.

\begin{figure}
\includegraphics[scale=0.30,angle=0]{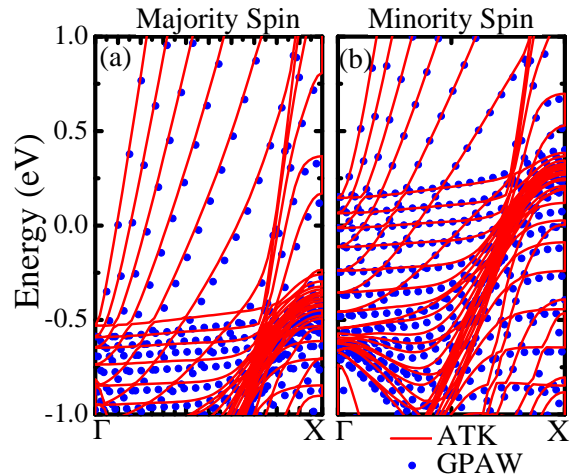}
\caption{The comparison of first-principles computed band structure of a periodic $\cdots$Ni$|$Gr$_5|$Ni$|$Gr$_5\cdots$ superlattice [with AC bonding configuration for Gr on the Ni(111) surface] obtained using either real-space grid PAW method implemented via the GPAW code~\cite{Enkovaara2010} or SZP basis of localized orbitals on C atoms and DZP basis on Ni atoms, together with pseudopotentials, implemented via the  ATK code.~\cite{quantumwise}}
\label{fig:fig4}
\end{figure}

The  NEGF-DFT framework was originally developed~\cite{Taylor2001,Brandbyge2002} to treat quantum transport through small molecules attached to metallic electrodes. Its application to modeling of charge and spin transport in MTJs requires careful tuning of pseudopotentials and basis sets in order to obtain an accurate description of the band structure near the Fermi level which is particularly important for the studies of spin-polarized transport. For example, pseudopotentials and localized basis sets that reproduce the electronic structure of the F electrode and barrier material alone do not necessarily reproduce the electronic structure of the more complicated F-electrode$|$barrier interfaces.~\cite{Waldron2007}

In order to capture accurately the electronic and magnetic structure around interfaces, we first compute the band structure of a periodic superlattice $\cdots$Ni$|$Gr$_5|$Ni$|$Gr$_5 \cdots$ using DFT based on the projector augmented wave (PAW) methodology and wave function representation on uniform real-space grids as implemented in the GPAW code~\cite{Enkovaara2010} where we choose grid spacing of \mbox{$0.18$ \AA}. Within the PAW formalism one works implicitly with the all-electron wave functions and has access to the (frozen) core states which make the method applicable to a broad range of systems (including materials with strongly localized $d$ or $f$ electrons that can be problematic to describe with pseudopotentials). The same band structure is then recomputed using DFT where wavefunctions are represented in terms of the linear combination of atomic orbitals (LCAO) and the behavior of the core electrons is described by norm-conserving Troullier-Martins pseudopotentials, as implemented in the ATK code.~\cite{quantumwise} In the ATK-based calculations, we choose single $\zeta$-polarized (SZP) basis on C atoms and double $\zeta$-polarized (DZP) basis on Ni atoms. The Brioullin zone of the superlattice was sampled by $12 \times 12 \times 100$ $k$-point grid, and the charge density and potentials were determined on a real-space grid with a mesh cutoff energy of 150 Ry. This was sufficient to achieve a total energy convergence of better than 0.01 meV/unit cell in the self-consistent loop.

The excellent agreement we achieve in Fig.~\ref{fig:fig4} between the real-space grid PAW and LCAO pseudopotential DFT calculations, where Perdew-Burke-Ernzerhof (PBE) parameterization of the spin-polarized generalized gradient approximation (GGA) for exchange-correlation functional has been used in both cases, also selects the correct parameters to be used for LCAO pseudopotential part of NEGF-DFT analysis of two-terminal Ni$|$Gr$_n|$Ni junctions discussed below. The  active region of the device in Fig.~\ref{fig:fig1}(a) simulated by the NEGF-DFT code consists of 7 Ni(111)  layers on the left, $n$ layers of Gr, and 6 layers of Ni(111) on the right. This active region is first relaxed until the maximum force component goes below $0.02 \ \mathrm{eV/\AA}$ per atom, and then attached to two semi-infinite ideal Ni electrodes.

The NEGF formalism for steady-state transport operates with two central quantities---the retarded ${\bf G}(E)$ and the lesser Green functions ${\bf G}^<(E)$---which describe the density of available quantum states and how electrons occupy those states, respectively. In the NEGF-DFT framework,~\cite{Areshkin2010,Taylor2001,Brandbyge2002} the Hamiltonian is not known in advance and has to be computed by finding the converged spatial profile of charge via the self-consistent DFT loop for the density matrix \mbox{${\bm \rho} = \frac{1}{2 \pi i} \int dE\, {\bf G}^<(E)$} whose diagonal elements give charge density.~\cite{Areshkin2010} The Hamiltonian matrix ${\bf H}$ in the local orbital basis $\{ \phi_i \}$  is composed of elements \mbox{$H_{ij} = \langle \phi_i |\hat{H}_{\rm KS}| \phi_{j} \rangle$}, where $\hat{H}_{\rm KS}$ is the effective Kohn-Sham Hamiltonian obtained from the DFT self-consistent loop and the overlap matrix ${\bf S}$ has elements \mbox{$S_{ij} = \langle \phi_i | \phi_j \rangle$}.

In the coherent (i.e., in the absence of electron-phonon or electron-electron dephasing processes) transport regime, only the retarded Green function
\begin{equation}\label{eq:retardedgf}
\mbox{${\bf G}^\sigma_{{\bf k}_\parallel}=[E{\bf S} - {\bf H}^\sigma_{{\bf k}_\parallel} - {\bm \Sigma}_{L,{\bf k}_\parallel}^\sigma - {\bm \Sigma}_{R,{\bf k}_\parallel}^\sigma]^{-1}$},
\end{equation}
of the active device region is required to post-process the result of the DFT loop by expressing the current between the left (L) and the right (R) electrodes
\begin{equation}\label{eq:current}
I^\sigma(V_b) = \int_{\rm BZ} \!\! d{\bf k}_\parallel \!\! \int \! dE \, T^\sigma({\bf k}_\parallel,E,V_b) [f(E-\mu_L)-f(E-\mu_R)].
\end{equation}
The electrodes are assumed to be attached to macroscopic reservoirs at infinity characterized by the Fermi function $f(E-\mu_{L,R})$, so that bias voltage driving the nonequilibrium transport is given by $\mu_L - \mu_R = eV_b$. Here we resolve all quantities in minority and majority spin channels---$\sigma = \mathrm{min}, \, \mathrm{maj}$---assuming irrelevance of spin-orbit coupling or spin-flip scattering. The spin- and ${\bf k}_\parallel$-resolved transmission function of coherent transport is given by
\begin{equation}\label{eq:transmission}
T^\sigma({\bf k}_\parallel,E,V_b) = {\rm Tr} \left\{ {\bm \Gamma}_{R,{\bf k}_\parallel}^\sigma (E)  {\bf G}^\sigma_{{\bf k}_\parallel}(E) {\bm \Gamma}_{L,{\bf k}_\parallel}^\sigma (E)  {\bf G}^{\sigma,\dagger}_{{\bf k}_\parallel}(E) \right\},
\end{equation}
where the level broadening matrices \mbox{${\bm \Gamma}_{L(R),{\bf k}_\parallel}^\sigma(E)=i[{\bm \Sigma}_{L(R),{\bf k}_\parallel}^\sigma(E) - {\bm \Sigma}_{L(R),{\bf k}_\parallel}^{\sigma,\dagger}(E)]$} are expressed in terms of the retarded self-energies ${\bm \Sigma}_{L(R),{\bf k}_\parallel}^\sigma(E)$ of semi-infinite ideal Ni electrodes. In order to converge integration over the (conserved in the absence of disorder) transverse wavevector ${\bf k}_\parallel$ in Eq.~\eqref{eq:current}, we find it necessary to use a dense grid $301 \times 301$ of $k$-points in the corresponding 2D BZ. This procedure yields the bias-dependent transmission function \mbox{$T(E,V_b) = \int_{\rm BZ} d{\bf k}_\parallel T({\bf k}_\parallel,E,V_b)$} plotted in Figs.~\ref{fig:fig3}(a) and \ref{fig:fig3}(b).

\section{Magnetoresistance at finite bias voltage}\label{sec:mr}

The Ni$|$Gr$_n|$Ni multilayered heterostructure is not a conventional MTJ. Unlike MgO-based MTJs where linear-response ($V_b \rightarrow 0$) conductances $G_{\rm P}^{\rm min}=I^{\rm min}/V_b$ and $G_{\rm P}^{\rm maj}=I^{\rm maj}/V_b$ decay exponentially~\cite{Butler2001} with increasing number of MgO layers, in the case of Ni$|$Gr$_n|$Ni junction $G_{\rm P}^{\rm min}$ is independent of $n$ for $n>4$ (apart from an even-odd oscillation as a function of the thickness $n$).~\cite{Karpan2007,Karpan2008} On the other hand, Gr$_n$ acts as a tunnel barrier for majority spin electrons causing $G_{\rm P}^{\rm maj}$ to decay exponentially with $n$. The spin-resolved linear-response conductances for Ni$|$Gr$_5|$Ni junctions are compared in \ref{tab:g} with the same conductances~\cite{Butler2001} for Fe$|$MgO$|$Fe MTJ containing MgO barrier of similar thickness as our Gr$_5$ barrier.

The recent first-principles analysis~\cite{Kuroda2011} of different metal$|$Gr$_n|$metal junctions for $n \le 4$, assuming reasonable metal-graphene epitaxial relationships, has delineated conditions for Gr$_n$ to behave effectively as a tunnel barrier causing exponential decay of the conductance with increasing $n$ which requires crystal momentum mismatch between the bulk Fermi-level states in the metallic electrode and those in the Gr$_n$ barrier. Furthermore, the recent experiments~\cite{Lee2011} measuring {\em I-V} characteristics of metal$|$Gr$_n|$metal vertical junctions (with Ti/Pt used as top and bottom metal electrodes) has demonstrated application of the bias voltage up to \mbox{$|V_b| \le 1$ V} without encountering catastrophic breakdown while showing transitions from Ohmic $I \propto V_b$ (at very low bias) to power law $I \propto V_b^m$ ($m>2$) characteristics.

\begin{table}
\begin{center}
\begin{tabular}{c|ccc}
\hline
\hline
 & $G_{\rm P}^{\rm maj}$ & $G_{\rm P}^{\rm min}$ & $G_{\rm AP}^{\sigma}$ \\
\hline
Ni$|$Gr$_5|$Ni(111) AC & $0.43 \times 10^{-1}$ & $5.7$ & $0.12  \times 10^{-1}$ \\
Ni$|$Gr$_5|$Ni(111) AB & $0.18 \times 10^{-2}$ & $3.0$ & $0.22  \times 10^{-2}$ \\
Fe$|$MgO$|$Fe(100) & $0.8 \times 10^{-1}$ & $0.9 \times 10^{-3}$ & $0.85  \times 10^{-3}$ \\
\hline
\hline
\end{tabular}
\caption{\label{tab:g} The approximative values for the linear-response conductances, in units of $\Omega^{-1}(\mu\mathrm{m})^{-2}$, for Ni$|$Gr$_5|$Ni junctions in AC [Fig.~\ref{fig:fig1}(c)] and AB [Fig.~\ref{fig:fig1}(b)] bonding configuration for Gr on the Ni(111) surface and for P and AP orientations of the Ni magnetizations. The third row shows the same conductances computed in Ref.~\onlinecite{Butler2001} for Fe$|$MgO$|$Fe(100) MTJ containing six-layer MgO barrier.}
\end{center}
\end{table}

\begin{figure*}
\includegraphics[scale=0.66,angle=0]{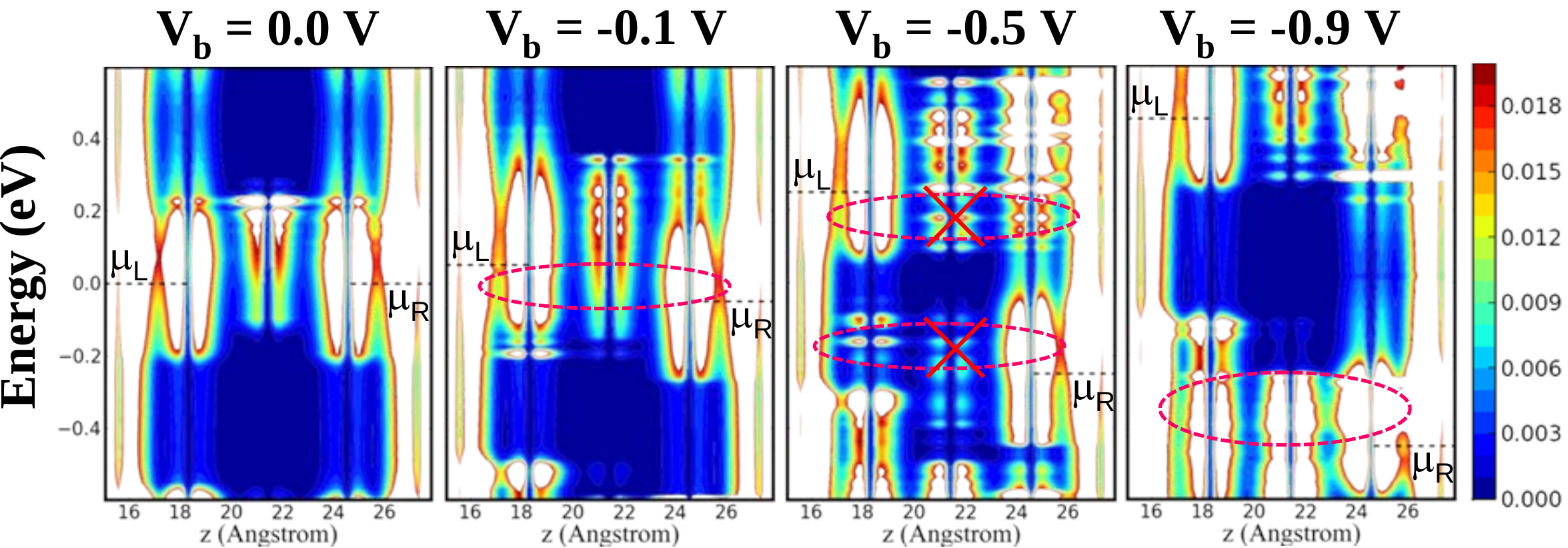}
\caption{The position-dependent LDOS from left to right electrodes in Ni$|$Gr$_5|$Ni junction, in AC bonding configuration at the Ni(111)$|$Gr interface and P orientation of the Ni magnetizations, at different bias voltages $V_b$. The electrochemical potentials $\mu_L$ and $\mu_R$ of the two Ni electrodes are marked by dashed horizontal lines while the zero of energy is set at ($\mu_L$ + $\mu_R$)/2. The LDOS exhibits high values in the Ni electrodes (white regions), while the central colored region corresponds to the Gr$_5$ barrier. The dashed ovals indicate the position of the resonant states which contribute to transport. Note that a strong coupling of the resonant states of the electrodes and the Gr barrier at a given energy level is required for large transmission $T(E,V_b)$ through the junction.}
\label{fig:fig5}
\end{figure*}

In conventional MTJs, tunneling rates are higher if there are similar or identical states on both sides of the barrier. Therefore, the tunneling electrons need not only to get through the barrier but there must be a state of the correct symmetry on the other side to accept them.~\cite{Butler2001} This effect is part of the reason for the commonly observed decrease in TMR with $V_b$ since as the bias increases the states on opposite sides of the barrier for P orientation differ more.~\cite{Waldron2007,Rungger2009,Zhang2004b} On the other hand, ``pessimistic'' MR in Ni$|$Gr$_n|$Ni remains 100\% up to \mbox{$V_b \lesssim 0.4$ V} for Gr barrier of thickness $n \ge 5$, as shown in Fig.~\ref{fig:fig2}(b).

\section{Negative differential resistance}\label{sec:ndr}

Figures~\ref{fig:fig3}(c) and ~\ref{fig:fig3}(d) plot the {\em I-V} characteristics for P and AP orientations of the Ni magnetizations where the total charge current is $I=I^{\rm min}+I^{\rm maj}$. Since in the AP orientation the bias-dependent transmission  $T(E,V_b)$ in Fig.~\ref{fig:fig3}(b) is nearly flat around the Fermi level, the {\em I-V} characteristics in Fig.~\ref{fig:fig3}(d) is linear up to the voltage \mbox{$V_b \approx \pm 0.6$ V}. However, in P orientation the total charge current $I$ sharply increases to reach its maximum value at \mbox{$V_b \approx \pm 0.12$ V} and then drops, thereby exhibiting a pronounced NDR. This feature can be explained using $T(E,V_b)$ curves plotted for AC configuration in Fig.~\ref{fig:fig3}(a). At lower $V_b$, the transmission resonance (around  \mbox{$E-E_F = 0.0$ eV}) falling into the bias window (marked by dashed wedge) contributes to the  peak in the  {\em I-V} characteristics. However, this resonance gets diminished with increasing $V_b$ which eventually shuts off the  current flow when \mbox{$V_b \approx 0.5$ V} is reached. The current is allowed to flow again when the new resonance around $E-E_F \approx -0.45$ eV enters the bias window \mbox{$V_b \approx -0.7$ V}.

Further insight into the microscopic mechanism behind NDR in P orientation of magnetizations in Ni$|$Gr$_n|$Ni junctions can be explained by examining the position-dependent LDOS
\begin{equation}\label{eq:ldos}
N(z,E)=-\frac{1}{\pi}\int d{\bf k}_\parallel \sum_{ij,\sigma} {\mbox{Im}} \langle \phi_i(z)| {\bf G}_{{ij},{\bf k}_\parallel}^\sigma(E)|\phi_j(z) \rangle,
\end{equation}
from the left to the right Ni electrode. The LDOS is plotted in Fig.~\ref{fig:fig5} where we choose four bias voltage values ({$V_b = 0.0, -0.1, -0.5, -0.9$ V) at which the magnitude of the total charge current differs significantly. In  equilibrium ($V_b = 0$ V), a prominent resonant state (white and red region) in the central Gr$_5$ region is located close to the Fermi level and couples well with the conduction states on both sides of Ni electrode. Upon application of the bias voltage, both the position and the width of resonant states start to change. At \mbox{$V_b = -0.1$ V}, part of the resonant state enclosed by the dotted oval still follows rigidly the upward-moving conduction state in the left Ni electrode while extending all the way to the downward moving conduction state in the right Ni electrode. As a result of this strong coupling between resonant conduction states within the energy interval $[\mu_L,\mu_R]$ enclosed by the electrochemical potentials of the two electrodes, the current increases notably. However, the charge density between the electrodes rapidly gets modified with the application of higher bias, and at $V_b = -0.5$ V, the resonant state splits into two parts (at energies $\pm 0.2$ eV) thereby losing coupling to one of the Ni electrodes. Thus, almost no current flows at this bias voltage. Increasing $V_b$ further introduces a new state in the central region at energy $\approx -0.4$ eV which couples strongly to both Ni electrodes at \mbox{$V_b = -0.9$ V} so that current starts increasing again.

\section{Concluding remarks}\label{sec:conclusion}

In conclusion, we demonstrated that perfect spin filtering in Ni$|$Gr$_n|$Ni, with $n \ge 5$ layers of graphene sandwiched between two (111) fcc Ni electrodes, characterized by ``pessimistic'' TMR=100\% at zero bias voltage~\cite{Karpan2007,Karpan2008} would persist even at finite bias voltage \mbox{$V_b \lesssim 0.4$ V}. This feature is markedly different from conventional MgO-based MTJs where TMR drops sharply~\cite{Waldron2007,Rungger2009} with increasing bias voltage. Thus, it could play an important role in spintronic devices based on STT.~\cite{Katine2008,Wang2011} Furthermore, we predict that Ni$|$Gr$_n|$Ni junction with P orientation of the Ni magnetizations would exhibit negative differential resistance as the bias voltage is increased from $V_b=0$ V to $V_b \simeq 0.5$ V due to transmission resonance formed at zero bias voltage which is then gradually pushed outside of the bias window.

\begin{acknowledgments}
We thank K. S. Novoselov and J. Q. Xiao for illuminating discussions. This work was supported by DOE Grant No. DE-FG02-07ER46374 (K. K. S. and B. K. N.) and Danish National Research Foundation's Center for Nanostructured Graphene (K. S. T.). The supercomputing time was provided in part by the NSF through XSEDE resource TACC Ranger under Grant No. TG-DMR100002 and NSF Grant No. CNS-0958512.
\end{acknowledgments}



\end{document}